\documentclass[]{spie}  

\usepackage{amsmath,amsfonts,amssymb}
\usepackage{graphicx}
\usepackage{newtxtext,newtxmath}
\usepackage{booktabs}
\usepackage[T1]{fontenc}
\usepackage{ae,aecompl}
\usepackage{multirow}
\usepackage{placeins}
\usepackage{adjustbox}
\usepackage{multirow}
\usepackage{adjustbox}
\usepackage[colorlinks=true, allcolors=blue]{hyperref}

\title{Towards operational optical turbulence forecast systems at different time scales}

\author[a]{Elena Masciadri}
\author[a]{Alessio Turchi}
\author[a]{Luca Fini}
\affil[a]{INAF-Osservatorio Astrofisico di Arcetri, L.go Enrico Fermi 5, 50125 Firenze, Italy}

\authorinfo{E-mail: elena.masciadri@inaf.it}
\begin{document} 
\maketitle

\begin{abstract}
The forecast on a time scale of 1 or 2 hours is crucial for all kind of new generation facilities (ELTs) instrumentation supported by the adaptive optics that will be mainly operated in Service Mode. In a recent study (Masciadri et al. 2020) we have showed that we can forecast the seeing and atmospheric parameters at such short time scales using an autoregressive method achieving unprecedented model accuracies with a substantial gain with respect to forecasts performed the day before (i.e. on longer time scales) obtained with an atmospherical mesoscale model. Equally we showed a gain with respect to the method by persistence using simply real-time measurements in situ on the same short time scale (1-2 hours). The auto-regressive method makes use of the forecasts done with mesoscale atmospherical models and real-time measurements and since 2019 has been implemented in the operational forecast system ALTA Center supporting LBT observations. In this contribution we apply the same approach to the VLT site extending the method to all the main astroclimatic parameters i.e. the seeing, the wavefront coherence time, the isoplanatic angle and the ground layer fraction. We prove that such a method offers unprecedented forecast accuracies for all the astroclimatic parameters with clear gains with respect to the prediction by persistence. Preliminary calculations indicate also better accuracies than those obtained with the machine learning based approach using in situ measurements.  We will apply soon such a method in an operational forecast system that we conceived for the VLT. 
\end{abstract}

\keywords{optical turbulence - atmospheric effects - numerical modelling - adaptive optics}

\section{INTRODUCTION}
\label{sec:intro}  

The classical and typical questions that our community has been asking for in the last decades concerning the optical turbulence forecast have been: (1) Are we able to forecast the optical turbulence (C$_{N}^{2}$ profiles and integrated astroclimatic parameters) ? (2) Which atmospheric models can we use ? (3) Which performances can we achieve ?

At present we can say that it has been shown/proven that it is possible to forecast the optical turbulence. Different typologies of atmospherical models have been used such as mesoscale models and General Circulation Models (GCM) even if the most suitable ones revealed to be the non-hydrostatical mesoscale models, mainly thanks to a higher horizontal resolution that is suitable to be sub-kilometric \cite{masciadri1999}. 

The methods employed for the OT forecast are mainly two: \\
(a) numerical approach: in which the C$_N^2$ is parameterized using the prognostic turbulent kinetic energy equation \cite{masciadri1999,masciadri2001,masciadri2002,masciadri2004b,masciadri2006,cherubini2008,lascaux2009,lascaux2010,hagelin2011,lascaux2011,masciadri2013,masciadri2017,masciadri2020,basu2020,lyman2020}. \\     
(b) analytical approach: in which the C$_N^2$ is expressed as a function of the temperature and the wind speed using algorithms mainly obtained with empirical fits or physical considerations \cite{dewan1993,vanzandt1978,trinquet2007,ruggiero2002,ye2011,giordano2013,osborn2018,wu2020}.\\

So far not too much attention has been given to the definition of the forecast time scale. This represents a great limitation (1) because we can not appreciate the goodness of a forecast; (2) because we can not compare an approach with another one; (3) because we can even risk to incur on errors in comparing things that can not be compared (it is to so rare in the literature) as it has been shown by Masciadri et al. \cite{masciadri2019}. 

The forecast is mainly a problem of time scale. 
There are many different typology of time and intervals of time that play a role during a forecast. 
We have, for example, the time in which the initialisation data are calculated, the time in which the simulation starts, the time in which a forecast is displayed therefore therefore the information arrives to us. We have also intervals of times such as the simulated time (ex: we can simulate the coming night starting from the beginning of the night and the end of the night); we have the time required to perform the forecast that depends on the hardware
and finally we have what we can call 
"forecast time scale" that is the interval between the time at which the information arrives to me and the time at which the information refers to. If, for example, we display the forecast of the coming night at 14 LT, the forecast time scale for the coming night starting at T$_{ini}$ and ending at T$_{end}$ is between around 6h and 15h. We call this 'standard configuration' as this is the usual configuration that we are using since some years to forecast the optical turbulence above the astronomical sites. 

The question are therefore: (1) can we perform forecasts at shorter time scale (1h or 2h) ? (2) can we provide operational forecasts at different time scales (with associated different accuracies) ? 

Time scales of 1h and 2h are the most critical ones for the science operation therefore the questions are challenging but extremely timely because observatories are more and more demanding with respect to the science operation.

The method we proposed to achieve these goals is based on a autoregression filter. Details of the method can be found in \cite{masciadri2020}. Here we remember the most fundamental elements. The forecast at the time (t+1) is given by the forecast at the time (t+1) of the atmospherical model at mesoscale plus a function of the difference between the atmopsherical model and observations calculated at the same time (t+1) and depending on a finite number of coefficients called regressors. The observations refer to real-time measurements related to a finite number of nights in the past. With this in mind the observational configuration is transformed in the following: at 14 LT, as usual, we provide a forecast of the parameter we are interested on for the coming night. When the night starts i.e. when real-time measurements start to be available, at each full hour we perform the forecast with the AR method on the successive four hours of the same parameter. If we repeat this procedure at  each full hour we obtain a forecast on a time scale of 1h (see Fig.2\cite{masciadri2020}). 

\begin{figure}
\begin{center}
\includegraphics[width=0.6\textwidth]{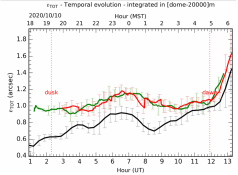}
\end{center}
\caption{\label{fig:seeing_ar} Temporal evolution of the seeing obtained using the standard configuration approach (long-time scale) and the AR method on a time scale of 1h (short-time scale).}
\end{figure}

On Fig.\ref{fig:seeing_ar} is shown the example of the temporal evolution during one night of the seeing forecasted with the standard configuration (black line) and obtained with the AR method on a time scale of 1h (red line). It is evident as the second one is much better correlated to real-time measurements (green line than the first one. 
In Section \ref{model} and Section \ref{obs} we summarise the main information related the model configuration used and the observations used to quantify the forecasts accuracy.

\section{MODEL}
\label{model}

In this contribution we used the non-hydrostatical mesoscale \cite{lac2018,lafore1998} and the ASTRO-MESO-NH package \cite{masciadri1999} developed for the optical turbulence (i.e. C$_N^2$ and integrated astroclimatic parameters). We used the optical turbulence parameterisation that allow a more suitable temporal variability of the turbulence in the free atmosphere \cite{masciadri2017}.
The geographic coordinates of Cerro Paranal  are (24$^{\circ}$37'33.117" S, 70$^{\circ}$24'11.642 W) and the height of the summit is 2635~m above the sea level. We used a grid-nesting technique \cite{stein2000} consisting in using different embedded domains of the digital elevation models (DEM i.e. orography) extended on smaller and smaller surfaces, with progressively higher horizontal resolution but with the same vertical grid. Simulations of the OT are performed on three embedded domains where the horizontal resolution of the innermost domain is $\Delta$X = 500~m. The model uses the \cite{gal1975} coordinates system on the vertical and the C-grid in the formulation of \cite{arakawa1976} for the spatial digitalization. In this study, we used in the wind advection scheme the 'forward-in-time' (FIT) numerical integrator instead of the' leap-frog' one. Such a solution allows for longer time steps and therefore shorter computing time. The model employs a one-dimensional 1.5 turbulence closure scheme \cite{cuxart2000} and we used the one-dimensional mixing length proposed by \cite{bougeault1989}. The surface exchanges are computed using the interaction soil biosphere atmosphere (ISBA) module \cite{noilhan1989}. 

The model is initialised with forecasts provided by the General Circulation Model (GCM) HRES of the European Center for Medium Weather Forecast (ECMWF) having an intrinsic horizontal resolution of around 9 km. For the short time forecasts obtained with the AR we used the same configuration used in \cite{masciadri2020}. We refer the reader to the paper \cite{masciadri2022} for further details on the model configuration used for the forecast delivery.

\section{OBSERVATIONS}
\label{obs}

Observations used for this study refer to the ESO archive, more precisely to the observations that are routinely done at the VLT to monitor the optical turbulence conditions during the night. ESO observations are stored in a public repository\footnote{{\href{http://archive.eso.org/cms/eso-data/ambient-conditions.html}{http://archive.eso.org/cms/eso-data/ambient-conditions.html}}} since April 2016. Observations we use refer to a Differential Image Motion Monitor (DIMM) \cite{sarazin1990} and a Multi Aperture Scintillation Sensor (MASS) \cite{kornilov2003,kornilov2014}. We treated also observations from a Stereo-
SCIDAR (SS) \cite{butterley2020} to quantify the measurements accuracy. The Stereo-SCIDAR is not a monitor but an instrument suitable for dedicated site testing campaigns as it requires a telescope with a diameter of at least 1 m. Said that a Stereo-SCIDAR runned for a few years at the VLT at the focus of an AT. ESO collected so far a rich statistical sample of measurements related to 157 nights (Release2019B).

\section{RESULTS}
\label{results}

\begin{figure*}
\begin{center}
\includegraphics[width=0.6\textwidth]{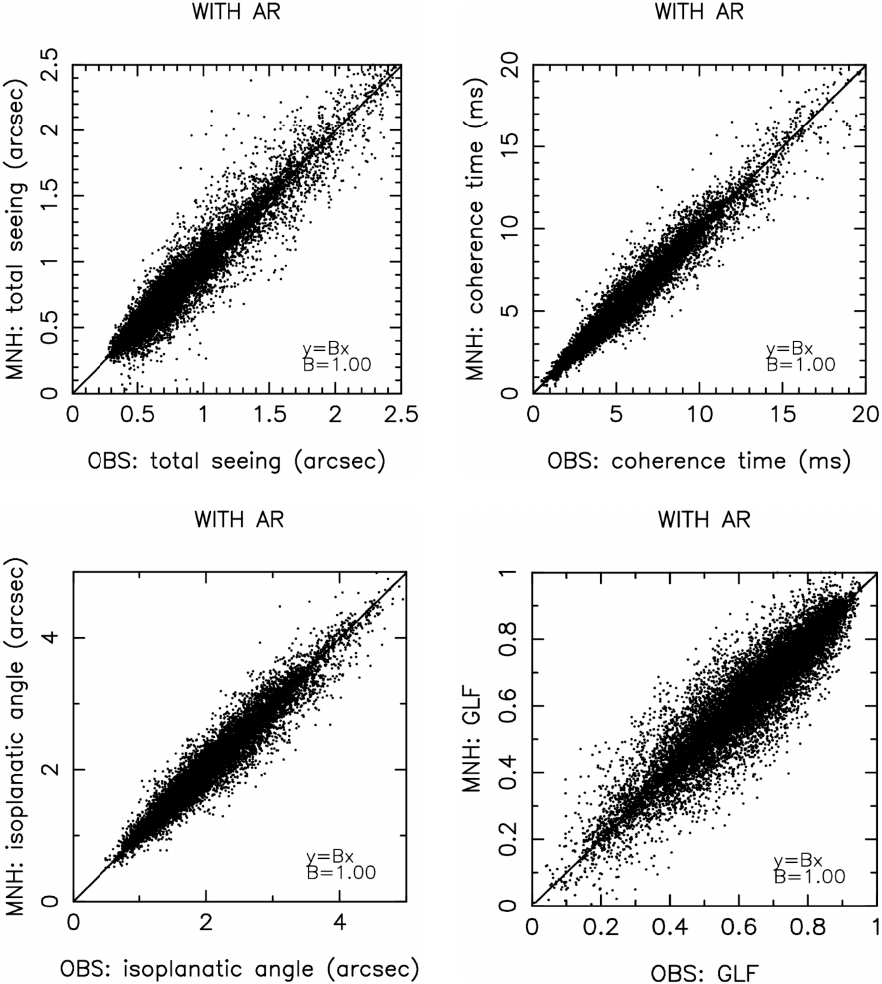}
\end{center}
\caption{\label{fig:scat} Scattering plot of observed and forecasted parameters using the AR method on a time scale of 1h: seeing ($\varepsilon$) (top-left), wavefront coherence time ($\tau_0$) (top-right), isoplanatic angle ($\theta_{0}$) (bottom-left) and GLF x 100 (\%) (bottom-right). B is the slope of the regression line. For the seeing observations are taken from the DIMM, for $\tau_{0}$, $\theta_{0}$ and GLF observations are taken from the MASS-DIMM. The density map of all the individual couple of points is shown in Fig.\ref{fig:C1} }
\end{figure*}

\begin{figure*}
\begin{center}
\includegraphics[width=0.7\textwidth]{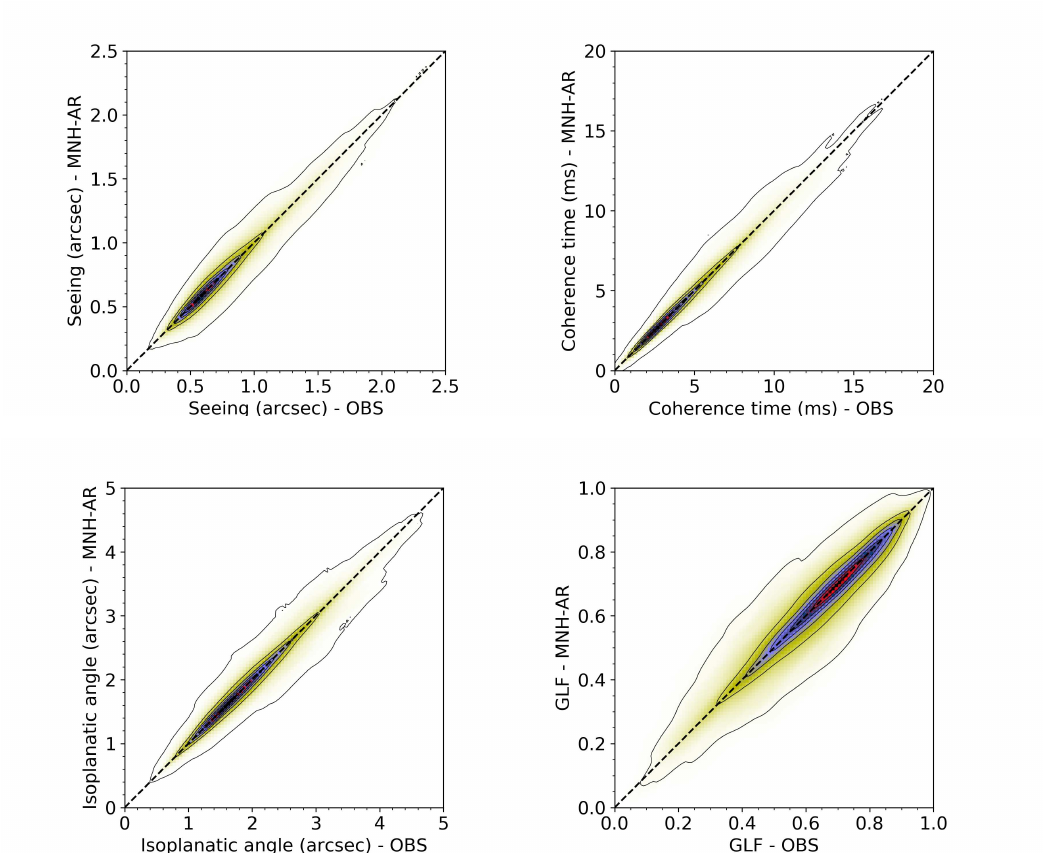}
\end{center}
\caption{\label{fig:C1} Density function map related to Fig.\ref{fig:scat}. }
\end{figure*}

To quantify the performances of the AR method we considered a statistical sample of 1 full solar year as we need a rich statistical sample with contiguous nights. Figure \ref{fig:scat} shows the scattering plots of forecasts versus observations for seeing, wavefront coherence time, isoplanatic angle and ground layer fraction. Figure \ref{fig:C1} shows the density function maps. From the latter is visible that the peaks of the distribution fall basically on the bisector giving therefore a null bias. To quantify the forecast performances we can follow different approaches. We report here just a preliminary analysis. A more extended analysis is presented on \cite{masciadri2022}.

We calculated the RMSE on each individual night, after we calculated the cumulative distribution on the whole sample of nights related to the statistical sample of 1 full solar year and then we retrieved the median value and the quantiles. Table \ref{tab:ind_med} reports the median value and the first and fourth quartiles calculated for all the astroclimatic parameters at a time scale of 1h and 2h. 

\begin{figure*}
\begin{center}
\includegraphics[width=0.8\textwidth]{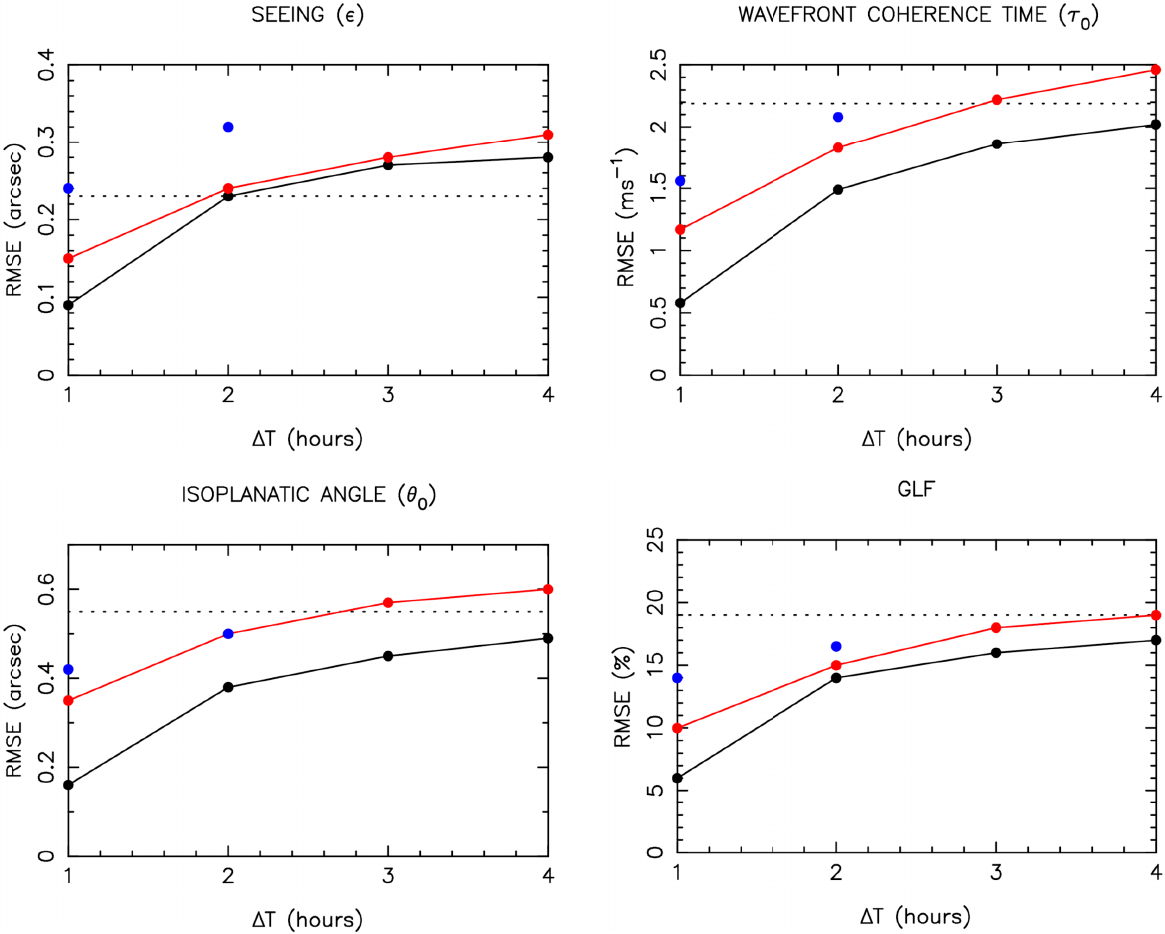}
\end{center}
\caption{\label{fig:mach_learn} RMSE vs different FTS $\Delta$T for seeing, wavefront coherence time, isoplanatic angle and GLF. On the x-axes the 'forecast time' $\Delta$T = (T$_{f}$ - T$_{i}$), where T$_{i}$ is the time in which the forecast is calculated and T$_{f}$ is the time that the forecast refers to. Ex: $\Delta$T = 1 means a forecast at 1h calculated at T$_{i}$. Black dashed line is the RMSE related to the standard forecast. Black line: AR method. Red line: forecast by persistent method. Blue dots: machine learning algorithm Random Forest (RF).}
\end{figure*}

Besides that, we calculated how the AR performances varies as a function of the time scale. Figure \ref{fig:mach_learn} shows the RMSE versus the time scale. The horizontal dashed line represents the RMSE related to the standard configuration, the black line the RMSE obtained using the AR method, the red line that obtained using the prediction by persistence, the blue dots the RMSE obtained with the machine learning Random Forest algorithm. The Prediction by Persistence (PP) assumes that the value of the parameter remains constant for the successive X hours starting from the present time. For the prediction by persistence we tested different configurations taking the as starting point a precise t$_{0}$ or a the last X minutes with respect to the time t$_{0}$ and the differences obtained on the RMSE are negligible. When the full lines intersect the horizontal dashed line that means that the gain of the black (AR method) or red line (PP method) with respect to the standard configuration disappears and it is null. In Figure \ref{fig:mach_learn} are reported also results obtained with a machine learning algorithm, the random forest (RF).

\begin{table}
\begin{center}
\caption{\label{tab:ind_med} Median, first and fourth quartiles of the RMSE calculated for different astroclimatic parameters on individual nights: seeing ($\varepsilon$), wavefront coherence time ($\tau_{0}$), isoplanatic angle ($\theta_{0}$) and ground layer fraction GLF. Seeing is calculated taking all values $\le$ 1.5 arcsec (selection done obviously on observations). Sub/superscripts refer to the first and fourth quartiles.}
\begin{tabular}{ccccc}
\hline \\
 RMSE    &  $\varepsilon$  &  $\tau_{0}$ & $\theta_{0}$  & GLF \\
          &     (arsec) &    (ms) &   (arcsec)  & ($\%$)  \\ 
\hline \\
 @ 1h   & $0.08^{0.10}_{0.06}$ & $0.44^{0.65}_{0.25}$ &  $0.14^{0.19}_{0.11}$ &  $6.00^{7.50}_{4.65}$  \\ \\
 @ 2h  & 0.19$^{0.26}_{0.14}$ &  1.01$^{1.65}_{0.60}$ &   0.32$^{0.44}_{0.24}$ &  13.72$^{17.00}_{11.20}$   \\ \\
\hline
\end{tabular}
\end{center}
\end{table}

As it has been demonstrated in \cite{masciadri2022} the accuracy obtained by the forecast system up to 2h is at present $\le$ of the standard deviation obtained for all the four astroclimatic parameters between different instruments i.e. the Stereo SCIDAR and the DIMM or the MASS-DIMM (depending on the parameter studied). This tells us that, even if a forecast system is by definition always in continuous improvement phase, such a system is already enough robust to provide a substantial support to the science operation.


We summarize here the main results obtained from Fig.\ref{fig:mach_learn}:\\
(1) the AR method provides better performances than the prediction by persistence   for all time scales and for all the astroclimatic parameters. \\
(2) The gain of the AR method with respect to the prediction by persistence at a time scale of 1h for all the astroclimatic parameters is in the range [1.7 - 2.2] depending on the parameter\\
(3) In the case of the seeing, for $\Delta$T > 2h the standard deviation configuration dominates on the AR method. That means that the gain of our forecasts are even more important:\\
- For $\Delta$T $\le$ 2h the AR method gains with respect to the PP\\
- For $\Delta$T > 2h the standard configuration gains with respect to the PP\\
\noindent
(4) The AR method provides better performances than the RF at 1h and 2h for all the astroclimatic parameters\\
(5) In most cases the RF provides even worse performances than the PP. That tells us that the atmospheric models play a fundamental role in the optimization of performances and we can not get rid of them.\\

Our next steps will see (1) the test of different machine learning algorithms and (2) the use of machine learning joint to atmospherical models. Preliminary results are helping us in definying several free parameters \cite{turchi2022b}

Results we have obtained changed our perspective. Research enters definitely into a new era of hybrid techniques made of numerical techniques (i.e. atmospherical models) plus statistical techniques. The latter include a set of various approaches such as the autoregression method, Kalman filter, Machine Learning, Deep Learning, artificial neural networking (ANN), etc.

\section{CONCLUSIONS}
\label{concl}

In this contribution we discussed about the possibility to realise forecasts at different time scales. We proposed a method to improve the forecast performances at short time scale. In this contribution we present preliminary results that are treated more extensively in the paper \cite{masciadri2022}. We listed here the main conclusions achieved by our study:

\noindent
(1) We presented the AR method and we showed that it provides performances at our knowledge never achieved before (at least in the astronomical context) on a time scale of 1h and 2h. Preliminary results on a time scale of 1h for the median RMSE are: 0.08 arcsec for the seeing, 0.44ms for the wavefront coherence time, 0.16 arcsec for the isoplanatic angle, 6 $\%$ for the GLF. The probability of detection for all the POD$_{i}$: POD$_{1}$, POD$_{2}$ and POD$_{3}$ on a time scale of 1h is equal to 99$\%$.\\
   
\noindent
(2) The gain of of the forecasts performances of AR method with a time scale of 1h (short time scale) with respect to the standard configuration (long time scale) is a factor 2.6 - 3.8 range for the astroclimatic parameters.\\

\noindent
(3) We implemented the AR technique in ALTA Center, conceived to support the science operation of the Large Binocular Telescope (LBT). ALTA Center is currently providing operational forecasts at two time scales at the Large Binocular Telescope: \\
- at long time scales (6h - 15h) with forecast available at 14 LT for the coming night \\
- at short time scale (1h, 2h) (from 1h to 4h with forecasts upgraded hourly)\\
At or knowledge ALTA Center is the unique site that is providing such kind of products.\\

\noindent
(4) The hybrid approach will be used in the FATE project applied to the VLT. This project won an ESO Call for Tender for the providing dedicated forecasts of astroclimatic parameters and atmospherical parameters relevant for the ground based astronomy.\\

\noindent       
(5) We have still interest in using the numerical approach alone (atmospherical models) and in working in improving performances of atmospherical model parameterisations as the forecasts  at long time scales strictly depend on this approach.\\

\noindent
In terms of perspectives:\\ \\
\noindent
(1) Results obtained with the AR technique opens to a new era in the field of the OT forecast in the astronomical context as it pushes us in using hybrid approaches, at least for short time scale forecasts. \\

\noindent
(2) We intend to investigate alternative techniques in the field of the machine learning / ANN / Kalman filter to verify if there is still space for improvements at
short time scales. Our benchmark of reference is the AR method that at present provided us the best results. \\

\noindent
(3) We envisage to forecast new parameters such as the sky-background \cite{turchi2020} and the PSF figures of merit. These parameters does not depend only on the atmopshere but also on the AO system and the software used to simulate their behavior \cite{turchi2022}.

\acknowledgments 
 
 The authors thanks the Meso-Nh users supporter team who constantly works to maintain the model by developing new packages in progressing model versions. Initialisation data come from the GCM HRES of the ECMWF. This study has been co-funded by the FRCF foundation through the 'Ricerca Scientifica e Tecnologica' action - N.45103 and by the EU Horizon 2020 research and innovation programme under the grant agreement N. 824135 (SOLARNET). Digital elevation model at high resolution has been obtained thanks to the Shuttle Radar Topography Mission (SRTM). The authors thanks Angel Otarola and ESO Santiago and Garching staff supporting this study. This study made use of the scikit-learn Python packages.

\bibliographystyle{spiebib} 

\end{document}